\documentclass[aps,prl,preprint,superscriptaddress,letterpaper]{revtex4}

\usepackage[utf8]{inputenc}

\usepackage{amsmath, amssymb}
\usepackage{tabularx,booktabs,array,dcolumn}
\usepackage{setspace}
\usepackage{vmargin}
\usepackage{graphicx,psfrag,subfigure}
\usepackage{xcolor}

\usepackage{marvosym}

\usepackage[all]{xy}

\usepackage{tikz}
\usetikzlibrary{calc}
\usetikzlibrary{patterns}

\usepackage{threeparttable}
\usepackage{rotating}
\usepackage{multirow}

\newcommand{\mx}[1]{\boldsymbol{#1}}

\newcommand{\mr}[1]{\mathrm{#1}}

\newcommand{\cm}{cm$^{-1}$}

\def\Eh{$\text{E}_\text{h}$}

\def\ra0{\mx{a}}

\def\som{Supplementary Material}

\def\Sgp{$\Sigma_\text{g}^+$}
\def\Sup{$\Sigma_\text{u}^+$}

\def\X1Sgp{X\ ^1\Sigma_\mr{g}^+}
\def\Sgp1{^1\Sigma_\mr{g}^+}

\def\Sup1{^1\Sigma_\mr{u}^+}
\def\b3Sup{b\ ^3\Sigma_\mr{u}^+}
\def\a3Sgp{a\ ^3\Sigma_\mr{g}^+}

\def\B1Sup{B\ ^1\Sigma_\mr{u}^+}
\def\Bp1Sup{B'\ ^1\Sigma_\mr{u}^+}
\def\Pu1{^1\Pi_\mr{u}}
\def\C1Pu{C\ ^1\Pi_\mr{u}}

\def\e3Sup{e\ ^3\Sigma_\mr{u}^+}
\def\c3Pup{c\ ^3\Pi_\mr{u}^+}

\def\Sp{S_\text{p}}
\def\Se{S_\text{e}}

\def\epsi{\varepsilon}

\begin{document}

\title{%
Precise computation of rovibronic resonances of molecular hydrogen: $EF~^1\Sigma_\text{g}^+$ inner-well rotational states \\[0.5cm]
}

\author{Dávid Ferenc and Edit Mátyus}
\email{matyus@chem.elte.hu}

\affiliation{Institute of Chemistry, ELTE, Eötvös Loránd University, Pázmány Péter sétány 1/A, Budapest, H-1117, Hungary}

\date{\today}

\begin{abstract}
\noindent %
Selected states of the $EF\ \Sgp1$ electronic manifold of the hydrogen molecule
are computed as resonances of the four-body problem.
Systematic improvement of the basis representation for the variational treatment
is achieved through an energy-tracking optimization procedure. 
The resulting non-relativistic energy is converged within a few nano Hartree, while the predissociative width
is found to be negligible at this level of accuracy.
The four-particle non-relativistic energies are appended with relativistic and 
quantum electrodynamics corrections which close the gap between the experimental observations 
and earlier theoretical work.
\end{abstract}

\maketitle
\noindent%
The lowest-energy rotational and vibrational states of the ground electronic state 
of the hydrogen molecule have received much attention over the past decade. We have witnessed
several orders of magnitude improvement in terms of accuracy and precision both on 
this experimental \cite{HoBeSaEiUbJuMe19,LiMe09,AlDrSaUbEi18,H2diss18} 
and theoretical \cite{WaYa18,PuKoCzPa19} frontier of molecular physics.

The hydrogen molecule has several electronically excited states,
many of them are very interesting by their own, \emph{e.g.,} the famous double-well features
caused by avoided crossings. The rovibronic level structure is dominated by
non-adiabatic interactions among the states, not yet fully, quantitatively
understood.
In the meanwhile, many of these electronically excited states have been measured experimentally
to high precision \cite{AlDrSaUbEi18,HoBeMe18}, and rovibronic states corresponding 
to electronic excitations, \emph{e.g.,} the $EF$ and $GK~\Sgp1$ manifolds, 
have been used in the excitation sequences resulting in the ultra-precise 
dissociation energy of the lowest rovibrational levels of 
the ground electronic state \cite{LiMe09,AlDrSaUbEi18,H2diss18}.
Therefore, it is not only for purely theoretical interest to aim for 
a better and more complete theoretical description of electronically excited states of the hydrogen molecule.
In particular, the entire dynamical range of molecular hydrogen, 
which has already been experimentally studied,
spans an about 130\ 000~\cm\ broad energetic and a 15~bohr broad structural (proton-proton separation)
range, 
and includes a very large number of sharp spectral transitions which can be measured
to high precision. For this reason, we think that 
the computation of a variety of these rovibronic states would offer an excellent testing
ground for the numerous small `effects' 
which have been identified during the study of the ground state of H$_2$, 
see for example, Ref.~\cite{PuKoCzPa19}.
The present work cannot aim for including all these effects at once, 
but we wish to provide a good starting point by significantly improving upon earlier theory 
for the selected states. 

\begin{figure}
  \includegraphics[scale=1.]{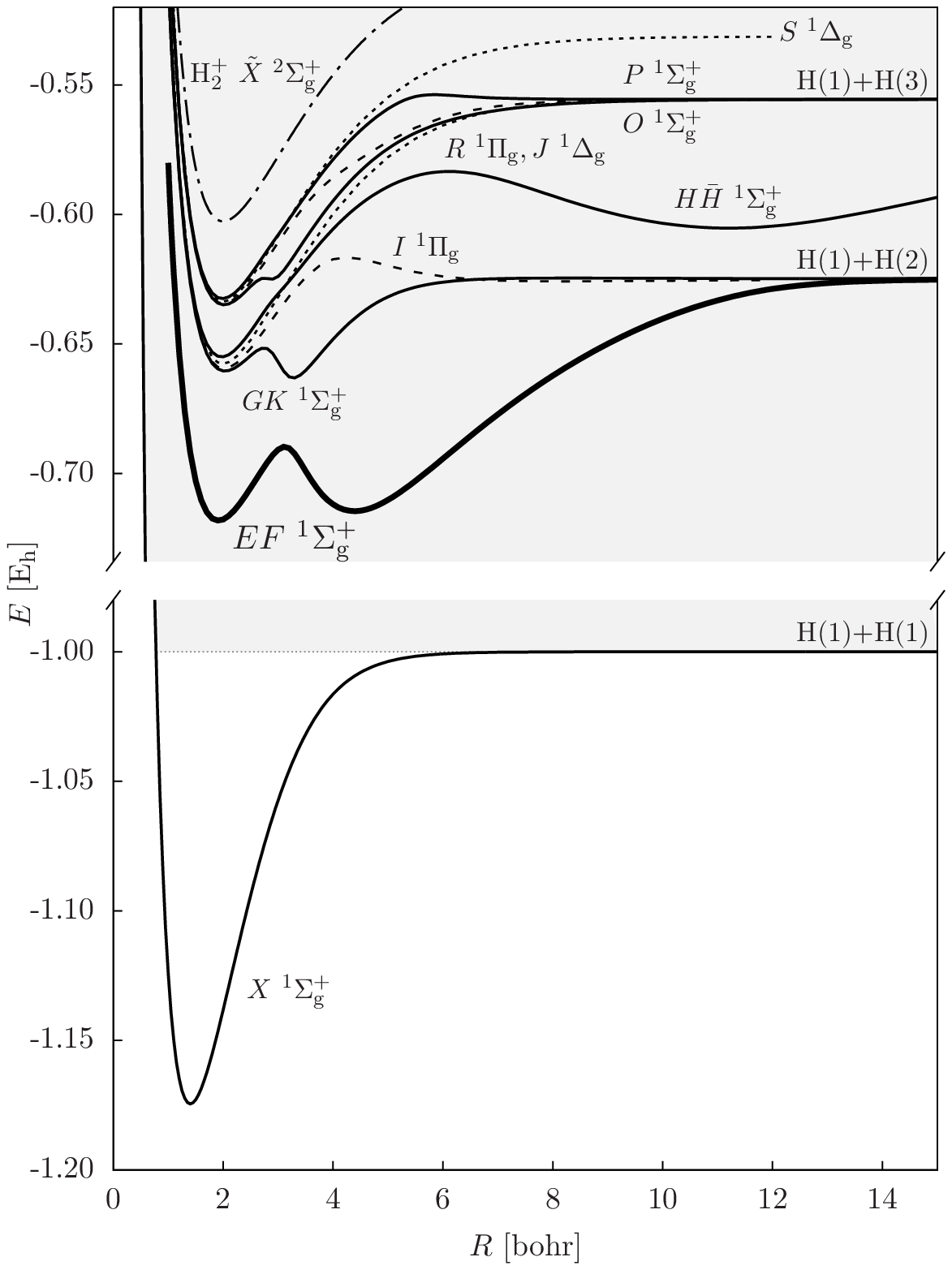}
  \caption{%
    Born--Oppenheimer potential energy curves for singlet gerade states of H$_2$ compiled from 
    Refs.~\cite{Wo95pi,Wo95delta,WoDr94,WoSiDa98}. The ground-state curve of H$_2^+$ \cite{BeMe16} 
    is also shown.
    The present work is concerned with rovibronic states
    which can be assigned to the $EF\ ^1\Sigma_\text{g}^+$ electronic state.
    \label{fig:bopes}
  }
\end{figure}

H$_2$ has several, challenging excited states, some of them 
are bound even in the four-body treatment, \emph{e.g.,}
the rovibronic levels of $\B1Sup$.
In the present work, we will look at the first electronically excited state beyond the ground state,
which is the $EF\ ^1\Sigma_\text{g}^+$ state. The Born--Oppenheimer (BO) potential energy
curve of $EF\ ^1\Sigma_\text{g}^+$ shows a double-well feature
due to an avoided crossing with the nearby $GK\ ^1\Sigma_\text{g}^+$ state (Figure~\ref{fig:bopes}).
In a pre-Born--Oppenheimer (pre-BO) description \cite{MaRe12,Ma13,Ma19review}, 
`all' non-adiabatic couplings and `effects' are automatically included, so we will not use
potential energy curves, nor coupling vectors in the computations, 
but the curves are useful to look at and we will
continue to use the electronic state labels to have a short description and reference 
for the computed four-particle states.
Since all non-adiabatic couplings are included, 
also the coupling with the $X\ ^1\Sigma_\text{g}^+$ (ground) state continuum is present (we cannot separate it), 
and thus the $EF$ states can only be obtained 
as resonances within the four-body problem \cite{Ma13}.
The lower-energy vibrations of $EF$ have been estimated to have a very long predissociative lifetime 
(much longer than their radiative lifetime) 
due to their very weak coupling to the dissociation continuum of the 
$X\ ^1\Sigma_\text{g}^+$ ground
electronic state \cite{QuDrWo90,Ma13}.

The non-adiabatic manifold, which includes also the $EF$ state, 
was computed by Yu and Dressler \cite{YuDr94} 
by explicitly coupling of nine electronic states.
Yu and Dressler used accurate potential energy curves and non-adiabatic coupling vectors,
and their nine-state computation resulted in rovibronic term values
within 0.1--20~\cm\ of experiment. This nine-state computation
was a significant improvement
upon the earlier five-state study
of Quadrelli, Dressler, and Wolniewicz \cite{QuDrWo90}, which showed a larger, 
1.3--120~\cm\ deviation from the experimental results.
As it was pointed out by Hölsch, Beyer, and Merkt recently \cite{HoBeMe18}, 
performing a non-adiabatic computation with more than nine fully coupled electronic states
for this system is not obvious (one would need to include many more electronic states and probably
also the interaction with the H$_2^+\ + \text{e}^-$ ionization continuum)
but extension of non-adiabatic perturbation theory could be possible.
The effective Hamiltonian for the quantum nuclear motion over coupled electronic states
which perturbatively accounts for the effect of the distant electronic states 
(not included in the fully coupled electronic band) has been recently formulated \cite{MaTe19}
following Refs.~\cite{Te03,PaSpTe07}
and its numerical application, by generalizing the computational 
approach developed for a single-state non-adiabatic Hamiltonian \cite{Ma18he2p,Ma18nonad,PrCeJeSz17,PaKo09},
should soon follow.

However, benchmark energies and wave functions 
are best obtained from the direct solution of the four-particle Schr\"odinger equation.
Some time ago \cite{Ma13}, one of us has identified rovibronic resonances in pre-BO computations, 
also from the $EF$ manifold, by using symmetry-adapted, explicitly correlated Gaussian (ECG) 
basis functions optimized for bound states. The global-vector representation \cite{SuVaBook98,SuUsVa98}
of ECGs made it possible to have a general basis set for $N\geq 0$ total angular momentum quantum number.
At that time, a systematic improvement scheme of the basis representation for resonances was not
available, which made it difficult to assess the accuracy of the results.
Although a complex analogue of the real variational principle exists for the complex-coordinate
rotated (CCR) Hamiltonian \cite{KuKrHo88}, the practical utilization of this complex variational principle
for basis function optimization is not straightforward (of course, 
its utility for computing the energy and the width for a given basis set is well established).
In this work, we propose a practical approach inspired 
by the stabilization method \cite{HaTa70,KuKrHo88,MaTaRyMo94}, 
which will be useful for the systematic improvement of the basis set for
long-lived resonance states that are only weakly coupled to the continuum. 

It is well known that basis functions can be optimized for the $n$th excited bound state
by minimizing the $n$th eigenvalue of the Hamiltonian matrix \cite{SuVaBook98,Ko18}.
If we used this procedure to minimize the energy of a resonance 
(by picking the $n$th eigenstate in a starting basis set, 
which is beyond the dissociation threshold and resembles
most the resonance to be computed) 
then, after a few basis refinement cycles, discrete
representations of the continuum would start to accumulate in our energy list 
and we would end up minimizing the energy converging to the dissociation threshold.
In order to avoid this minimization `collapse', we do not focus on the $n$th state
but we define an energy threshold, $\epsi$, the energy of which is slightly smaller than
the exact energy of the state we are looking for. (This value can be estimated and does not need
to be a very tight estimate. In practice, we use the experimental term to 
set a loose lower bound.) We use this `energy-tracking' procedure
to optimize the basis representation for the selected rovibronic state 
by minimizing the energy of the first state above the $\epsi$ energy threshold. 
It is important to add that the energy tracking (in its simplest form) works for 
long-lived resonances, which couple to the continuum only weakly,
and the energy of which in increasing but finite basis sets stabilizes 
over many iteration cycles. During the refinement cycles, discrete representations of the continuum 
may drop in our energy list, but we track and minimize the energy of that state which is the lowest above our
$\epsi$ threshold.
We use this procedure only to optimize the basis functions.
Following the basis optimization, we studied resonance features using 
the complex-coordinate rotation (CCR) technique \cite{KuKrHo88} and its
pre-BO implementation of Ref.~\cite{Ma13}.
The point at which the CCR trajectories stabilize in the complex energy plane defines 
the full (complex) molecular wave function of the resonance (the complex energy \text{and} the CCR angle).
This complex wave function can then be used to compute relativistic and quantum electrodynamics (QED) corrections,
using the CCR form of the correction operators, to obtain the relativistic and QED 
corrections for both the energy and the width (lifetime)
\cite{Ko04}.

Now, we focus our attention to the computation of
the rotational excitations, with $N=0,1,2,3,4,$ and $5$ total angular momentum
quantum numbers, of the lowest-energy vibration in the inner well
of the $EF\ ^1\Sigma_\text{g}^+$ electronic state.
The inner-well ground vibrational state was labelled with $E0$ in Ref.~\cite{YuDr94},
so we will use $E0N$ for the $N$th rotational excitation of this state.
We can access these states by choosing the appropriate non-relativistic quantum numbers,
which are listed in caption to Table~\ref{tab:numnr},
and by defining the $\epsi$ energy threshold for the energy-tracking procedure,
which we set to ca. 10--20~\cm\ lower than the energy estimated from the experimental term value.
Using the QUANTEN computer program \cite{MaRe12,Ma13,Ma19review} (see also the \som\ \cite{som}),
we optimized, in repeated refinement cycles, 
a starting basis set which was compiled 
from the extensive bound-state optimization work of Ref.~\cite{Ma13}. 
In retrospect, we can confirm that the 
largest basis set of Ref.~\cite{Ma13} gives $EF$ energies accurate
within ca.~10 nE$_\text{h}$, which we can estimate now from the convergence 
pattern and behaviour of the states upon the systematic refinement of the basis set.

In Table~\ref{tab:numnr}, we list the non-relativistic energy values optimized in the present work
($E^{(2)}$ column), which are estimated to be 
within a few times $1\ \text{nE}_\text{h}=10^{-9}\ \text{E}_\text{h}$ of 
the exact, non-relativistic value.
The non-relativistic term values, the difference of the $E0N$ energies and the non-relativistic energy of the ground 
state ($X00$), reduce the 0.3~\cm\ deviation of the nine-state non-adiabatic computation
of Yu and Dressler \cite{YuDr94} to 0.1~\cm, and
confirm their electronic energy error estimate.
Yu and Dressler also estimated the relativistic and QED corrections 
to be ca. 0.08~\cm\ for the $E0N$--$X00$ terms. 
This is an average value for the states with different rotational quantum numbers,
and was compiled from the expectation value of the Breit--Pauli Hamiltonian with the electronic wave function 
at $R=1.9$~bohr \cite{WoDr94}
and the QED correction of H$_2^+$ \cite{BuJeMoKo92} 
at the $R=1.9$~bohr proton-proton distance, which is
near the effective structure of the $E00$ state.
If we correct our non-relativistic term values with these estimates, then
the deviation from experiment reduces to 
$(0.035,0.036,0.040,0.045,0.050,0.057)$~\cm\
for $N=0,1,2,\ldots,5$, respectively.

We would like to have a more complete account for the relativistic and QED effects
and also to possibly know the $N$ dependence of the correction.
For this purpose, we compiled data from the literature \cite{Wo98EFrel} and carried out 
additional computations using the non-relativistic, pre-BO wave functions computed in the present work. 
The relativistic, leading and higher-order QED effects (we have explicitly considered estimates
up to the so-called $m\alpha^7$ terms) are calculated as 
perturbative corrections using the non-relativistic energy and wave function, $E^{(2)}$ and $\psi$,
in terms of increasing orders of the fine-structure constant, $\alpha$, 
\begin{align}
  E^{(2..7)}
  = 
  E^{(2)}
  +
  \sum_{k=2}^{5}
  \alpha^k \langle \psi |\mathcal{H}^{(k+2)}| \psi \rangle 
\end{align}
where the $\mathcal{H}^{(k+2)}$ operators are reproduced from the literature in the following paragraphs 
(with the usual meaning of the symbols and operators, the details can be found in the corresponding references).
The underlying integrals, to be described in the following paragraphs, 
are evaluated so that the uncertainty of each correction term is better than $0.001$~\cm.

To calculate the (spin-independent) relativistic correction, 
we have started out from 
the expectation value of the Breit--Pauli Hamiltonian (of the electrons) \cite{BeSabook75,PuKoPa17}
\begin{align}
  \mathcal{H}^{(4)}
  =
  &-\frac{1}{8} (\mx{p}^4_1+\mx{p}^4_2)
  +\frac{\pi}{2} 
  \sum_{i=1}^2 \sum_{a=3}^4 \delta (\mx{r}_{ia}) 
  \nonumber \\
  &+\pi \delta(\mx{r}_{12}) 
  - \frac{1}{2} 
  \left[%
    \mx{p}_1 \frac{1}{r_{12}} \mx{p}_2 
    + \mx{p}_1 \cdot \mx{r}_{12}\frac{1}{r_{12}^3}\mx{r}_{12} \cdot \mx{p}_2 
  \right] \; ,
  \label{eq:BP}
\end{align}
where the electrons are labelled with `1' and `2', while the protons are labelled with `3' and `4'.
Wolniewicz already calculated the expectation value of $\mathcal{H}^{(4)}$
with the electronic wave function 
along a series of nuclear separations \cite{Wo98EFrel,Woftp}. 
We obtained the relativistic correction to each $E0N$ state 
by evaluating the expectation value of the BO relativistic correction
curve (represented with polynomial fits) with the pre-BO wave functions.
To obtain the term corrections, $\delta T^{(4)}$ (Table~\ref{tab:numrelrad}),
we used the relativistic correction value, $-1.652$~\cm, of the $X00$ ground state 
derived from a similar level of theory \cite{PuKoPa17}. Note that this value is 0.002~\cm\
smaller than the correction calculated directly with 
the Breit--Pauli Hamiltonian of the electrons and protons and 
the four-particle wave function 
of the $X00$ state \cite{WaYa18},
which will have to be accounted for when we estimate the uncertainties of the present results.

The leading QED contribution (to the electronic part of the problem)
\cite{Ar57,Su58,BeSabook75,KoPiLaPrJePa11} 
is
\begin{align}
  \mathcal{H}^{(5)}
  =
  &
  \frac{4}{3}\left[%
    \frac{19}{30} - 2 \ln\alpha - \ln K
  \right]
  \sum_{i=1}^2
  \sum_{a=3}^4 \delta (\mx{r}_{ia})
  \nonumber \\
  &+
  \left[%
    \frac{164}{15}+ \frac{14}{3} \ln\alpha
  \right]
  \delta(\mx{r}_{12})
  -\frac{7}{6\pi} P(1/r_{12}^3)
  \label{eq:rad}
\end{align}
which we evaluated with accurate electronic wave functions
along a series of nuclear configurations.
The $\ln K$ non-relativistic Bethe-logarithm was also treated within the BO approximation
similarly to Ref.~\cite{PiJe09}. The $\ln K(R)$ values for the $EF$ electronic state 
were approximated with the $\ln K(R)$ function of the ion core of $EF$,
so we could use the accurate $\ln K(R)$ values of
(the lowest electronic state of) H$_2^+$ computed by Korobov \cite{KoHiKa13,Ko18}.
The Dirac delta terms containing the electron-proton and electron-electron displacement vectors were obtained 
similarly to the Darwin terms of the relativistic correction, \emph{i.e.,} by computing the expectation
value of the $R$-dependent correction curves with the pre-BO wave function for each $N$.
The last term in Eq.~(\ref{eq:rad}) is the Araki--Sucher (AS) correction, which 
we computed for the $EF$ state in the present work 
using accurate electronic wave functions 
(obtained within the BO module of QUANTEN using floating ECGs \cite{Ma18nonad,Ma18he2p})
and the integral transformation technique \cite{PaCeKo05}. 
The AS correction to each $E0N$ state was obtained as the expectation value
of the correction curve with the four-particle wave function,
and was found to be an order of magnitude smaller, $-0.001$~\cm, than the 
correction for the ground state, $-0.013$~\cm\ \cite{KoPiLaPrJePa11,PuKoCzPa19}.
By summing up all these contributions, we obtain
the leading QED correction to the energy, which changes from
$0.385$ to $0.379$~\cm\ as $N$ increases from 0 to 5. 
To calculate the $\delta T^{(5)}$ leading QED term corrections
listed in Table~\ref{tab:numrelrad}
we used the $0.736$~\cm\ value of the ground state 
compiled from Refs.~\cite{PuKoPa17,PuKoCzPa19}.

Higher-order QED corrections were estimated at the one-loop level by retaining only
those terms which give the dominant corrections at these orders \cite{PiJe09,PuKoCzPa16,PuKoCzPa19}
\begin{align}
  \mathcal{H}^{(6)}_{\text{est}}
  &=
  \pi\left(\frac{427}{96} - 2\ln 2\right)  
  \sum_{i=1}^2 \sum_{a=3}^4 \delta (\mx{r}_{ia}) 
  \label{}\\
  \mathcal{H}^{(7)}_{\text{est}}
  &=
  -4\ln^2 \alpha\ 
  \sum_{i=1}^2 \sum_{a=3}^4 \delta (\mx{r}_{ia}) . 
\end{align}
While the $\mathcal{H}^{(6)}_{\text{est}}$ contribution to the $E0N$--$X00$ term values is $-0.003$~\cm,
the $\mathcal{H}^{(7)}_{\text{est}}$ changes the terms by as little as $2\cdot 10^{-4}$~\cm, which 
is negligible given the uncertainty of the current evaluation of the relativistic and QED integrals.

The overall $\delta T^{(4..7)}$ contribution (Table~\ref{tab:numrelrad}) of 
the relativistic, leading and higher-order QED effects 
increases from 0.122 to 0.146~\cm\ upon the increase of $N=0$ to 5.
The resulting $T^{(2..7)}$ term values for $N=0,\ldots,5$
show $\pm 0.001$~\cm\ deviations from the experimental values of Ref.~\cite{DiSaNiJuRoUb12},
which is better (probably fortuitous) than the uncertainty of the present theoretical values, 
which are thought to be accurate within about $\pm 0.005$~\cm.

The experimental values of Ref.~\cite{DiSaNiJuRoUb12} are more precise than our theoretical results, 
and the additional significant digits surely hide interesting physics, 
so theory should aim for further improvements. 
In order to help future work, we close this article with commenting
on the possible sources of uncertainties in our work.

First of all, the non-relativistic energy was obtained in a variational procedure (stabilized for long-lived
resonances), and systematic improvement of this value is rather straightforward. 
The convergence pattern observed in repeated rounds
of refinement cycles suggests that the non-relativistic energy, $E^{(2)}$, is converged within $0.000\,5$~\cm.
Assessment of the uncertainty of the relativistic and QED corrections is more delicate.
In the light of the developments of recent years for the rovibronic ground state \cite{PuKoPa17,WaYa18,PuKoCzPa16,PuKoCzPa19}, 
we think that the largest source of error in our work must be due to the relativistic `recoil' effect, 
on the order of a few $10^{-3}$~\cm,
which in simple terms means that (at least) the relativistic corrections should be computed 
using the full electron-nucleus Breit--Pauli Hamiltonian and the four-particle wave functions 
\cite{WaYa18,PuKoCzPa19}.  
Then, in order to pinpoint one-two more digits in the calculations, 
the current approximations used for the non-relativistic Bethe-logarithm term will have to be checked
and the contribution of the neglected terms in the higher-order QED corrections 
(in particular, the full $m\alpha^6$ contribution
of $\mathcal{H}^{(6)}$) will have to be elaborated.

In the usual perturbative manner, relativistic quantum electrodynamics (and possibly beyond)
is adapted to molecular computations, it is necessary to evaluate and sum several, small (and often not so small) 
contributions (of different signs) on top of a direct, 
variational non-relativistic computation. We think that the extremely rich excited state, rovibronic level structure
of the hydrogen molecule (Figure~\ref{fig:bopes}) offers an excellent opportunity to challenge and cross-check
the theoretical and computational procedures
both in terms of the completeness of the physical description as well as regarding the error balance of possible uncertainties
and inaccuracies. The present work demonstrates that electronically excited states of H$_2$ can be 
theoretically described to high precision and, with further improvements, they will 
provide equally useful and at the same time complementary information 
to the study of the ground electronic state.

\vspace{0.3cm}
\paragraph*{Acknowledgement}
\noindent %
We acknowledge financial support from a PROMYS Grant (no. IZ11Z0\_166525)  
of the Swiss National Science Foundation.
DF thanks a doctoral scholarship from
the New National Excellence Program of 
the Ministry of Human Capacities of Hungary
(ÚNKP-18-3-II-ELTE-133). 
EM is thankful to ETH~Z\"urich for supporting a stay as visiting professor during 2019 and 
the Laboratory of Physical Chemistry for their hospitality, where
part of this work has been completed.

\clearpage


\clearpage

\clearpage
\begin{table}
  \caption{%
  Variational, non-relativistic four-particle energy, in $\text{E}_\text{h}$, 
  corresponding to the $N=0,1,...,5$ rotational states  of the ground vibrational state in the inner well, `$E0N$', 
  of the $EF$~$\Sgp1$ electronic state of H$_2=\lbrace \text{e}^-,\text{e}^-,\text{p}^+,\text{p}^+ \rbrace$.
  The term values, in cm$^{-1}$, are given with respect to the rovibronic ground state.
  To obtain these singlet ($\Se=0$) states, the parity and the proton spin were chosen 
  to be $p=(-1)^N$ and $\Sp=(1-p)/2$, respectively. 
  The $E^{(2)}$ non-relativistic energy is estimated to be converged within a few nano Hartree.
  \\
  \label{tab:numnr}
  }
  \begin{tabular}{@{}l@{\ \ \ }c@{\ \ \ }c@{\ \ \ }c@{\ \ \ }c @{}}
  \cline{1-5}\\[-0.4cm]
  \cline{1-5}\\[-0.4cm]  
  $N$ &  
  $E^{(2)}$\ &
  $T^{(2)}$\ $^\text{a}$ &  
  $\Delta T^{(2)}_\text{o-c}$ $^\text{b}$ &
  $\Delta T^{(2)}_\text{o-c}$ \cite{YuDr94} \\
  \cline{1-5}\\[-0.2cm]
  0 & $-$0.712\,197\,577 & 99164.664 & 0.123 & 0.320 \\
  1 & $-$0.711\,908\,569 & 99228.094 & 0.124 & 0.321 \\
  2 & $-$0.711\,332\,945 & 99354.429 & 0.128 & 0.304 \\
  3 & $-$0.710\,475\,421 & 99542.633 & 0.133 & 0.312 \\
  4 & $-$0.709\,342\,932 & 99791.186 & 0.138 & 0.32  \\
  5 & $-$0.707\,944\,454 &100098.116 & 0.145 & 0.33  \\
  \cline{1-5}\\[-0.4cm]
  \cline{1-5}
  \end{tabular}
  \begin{flushleft}
    $^\text{a}$~
    $T^{(2)}=E^{(2)}-E^{(2)}(X00)$, where the ground-state, non-relativistic energy
    is $E^{(2)}(X00)=-1.164 025 031$~\Eh\ \cite{PuKoCzPa19}. \\
    $^\text{b}$ 
    $\Delta T_\text{o-c}^{(2)}=T_\text{o}-T^{(2)}_\text{c}$ 
    deviation of the observed (o) and computed (c) term values, where
    $T_\text{o}$ is taken from Ref.~\cite{DiSaNiJuRoUb12}.
  \end{flushleft}
\end{table}

\begin{table}
  \caption{%
  Perturbative relativistic and QED corrections up to estimates for $m\alpha^7$ (see text), in \cm, to 
  the $E0N$--$X00$ term values of H$_2$ reported in Table~\ref{tab:numnr}.
  The relativistic and QED corrections are estimated to be accurate within $10^{-3}$~\cm, which 
  results an overall uncertainty estimate $\pm 0.005$~\cm\ for $T^{(2\ldots 7)}$.
  \label{tab:numrelrad}
  }
  \begin{tabular}{@{}l@{\ \ \ } c@{\ \ \ }c@{\ \ \ }c@{\ \ \ }c@{\ \ \ }c@{\ \ \ }c@{}}
  \cline{1-7}\\[-0.4cm]
  \cline{1-7}\\[-0.4cm]  
  $N$ &  
  $\delta T^{(4)}$ $^\text{a}$ & 
  $\delta T^{(5)}$ $^\text{b}$ &
  $\delta T^{(6..7)}$ $^\text{c}$ &
  $\delta T^{(4..7)}$ $^\text{d}$ &
  $T^{(2..7)}$ $^\text{e}$&
  $\Delta T^{(2..7)}_\text{o-c}$ $^\text{f}$   \\
  \cline{1-7}\\[-0.2cm]
  0 & 0.475 & $-0.351$ & $-0.0027$ & 0.122 & 99164.786 & $0.001$ \\
  1 & 0.478 & $-0.351$ & $-0.0027$ & 0.124 & 99228.217 & $0.001$ \\
  2 & 0.482 & $-0.352$ & $-0.0027$ & 0.127 & 99354.557 & $0.001$ \\
  3 & 0.488 & $-0.353$ & $-0.0027$ & 0.132 & 99542.764 & $0.000$  \\
  4 & 0.496 & $-0.355$ & $-0.0027$ & 0.138 & 99791.326 & $0.000$ \\
  5 & 0.506 & $-0.357$ & $-0.0028$ & 0.146 & 100098.265 & $-0.001$ \\
  \cline{1-7}\\[-0.4cm]
  \cline{1-7}\\[-0.4cm]    
  \end{tabular}  
  \begin{flushleft}
    $^\text{a}$
    Relativistic correction. \\
    $^\text{b}$   
    Leading QED correction. \\
    $^\text{c}$   
    $\delta T^{(6..7)}=\delta T^{(6)} + \delta T^{(7)}$ higher-order QED corrections
    estimated by the dominant contributions to the one-loop term. \\
    $^\text{d}$   
    $\delta T^{(4..7)}=\delta T^{(4)}+\delta T^{(5)}+\delta T^{(6)}+\delta T^{(7)}$. \\
    $^\text{e}$   
    $T^{(2..7)}=T^{(2)}+\delta T^{(4..7)}$. \\
    $^\text{f}$
    $\Delta T^{(2..7)}_\text{o-c}=T_\text{o}-T^{(4..7)}$ 
    deviation of the observed (o) and computed (c) term values, where
    $T_\text{o}$ is taken from Ref.~\cite{DiSaNiJuRoUb12}.
  \end{flushleft}
\end{table}

\end{document}